\documentstyle[12pt,epsf,epsfig]{article}    % Simple LaTeX, use with DINA4 as follows
\newcommand{\pomeron}{I\!\!P}
\newlength{\dinwidth}                       
\newlength{\dinmargin}                      
\def\lsim{\mathrel{\rlap{\lower4pt\hbox{\hskip1pt$\sim$}}
    \raise1pt\hbox{$<$}}}                % less than or approx. symbol
\def\gsim{\mathrel{\rlap{\lower4pt\hbox{\hskip1pt$\sim$}}
    \raise1pt\hbox{$>$}}}                % greater than or approx. symbol
\newcommand{\beq}  {\begin{equation}}
\newcommand{\eeq}  {\end{equation}}
\newcommand{\bmath}{\begin{eqnarray}}
\newcommand{\emath}{\end{eqnarray}}

\begin{document}
\begin{titlepage}
\begin{flushleft}
{\tt DESY 97-250    \hfill    ISSN 0418-9833} \\
{\tt December 1997}
\end{flushleft}
\vspace*{4.cm}

\begin{center}  \begin{Large} \begin{bf}
Topics on the Physics Potential and Detector Aspects\\ of 
the LC-HERA {\boldmath $ep$} Collider\\
  \end{bf}  \end{Large}

  \vspace*{2cm}
  \begin{large}
A.~De Roeck\\
 \end{large}
%\end{center}

\vspace*{.8cm}
Deutsches Elektronen-Synchroton DESY, Notkestr.\ 85, 22603 Hamburg,
 Germany\\

\vspace*{2.cm}
\begin{quotation}
\noindent
{\bf Abstract:}
We present a brief account of some 
 physics topics which could be addressed
at a  collider consisting of the HERA proton ring and the future
$e^+e^-$  Linear Collider. A few experimental aspects are also discussed.
\end{quotation}

\vspace{5.5cm}
\noindent

\begin{center}
{\it Invited talk at the}\\ {\it  ``Workshop on 
 LINAC-RING Type $ep$ and $\gamma p$ Colliders'', Ankara, April 1997.}
\end{center}

\vfill

\vfill
\cleardoublepage
\end{center}
\end{titlepage}
%%%
%%%   SECTION 1
\section{Introduction}

A  breakthrough in the study of deep inelastic lepton-hadron scattering has 
been achieved by HERA, which provides, for the first time,
collisions of electrons 
and protons in a collider mode, thereby increasing the centre of
mass system (CMS) energy of the scattering processes by an order of magnitude
compared to the traditional fixed target experiments.
Since  the maximal reachable energies for electrons in  circular
colliders are much lower than for protons, at HERA a 820 GeV proton beam 
collides with a 27.5 GeV electron beam.
A large increase in electron beam energy can only be expected from a 
Linear Collider (LC) type of machine.
 If a high energy electron-positron LC 
would be  build close to a laboratory
site where a high energy proton storage ring already exists, a new frontier
in the study of $ep$ scattering can be reached. The DESY laboratory in
Hamburg is studying the case to built
such a  LC~\cite{LC}. Present planning foresees the collider 
to be tangial to the existing proton ring of the HERA collider (HERAp), thus
allowing for  collisions of 250 (500) GeV electrons on 820 (1000) 
GeV  protons.
This would increase the CMS energy of the $ep$ collision by a factor 3-6
compared to HERA.
A study on the possible luminosity which could be reached by using HERA
in combination with the TESLA LC is given in~\cite{brinkman} and amounts to
roughly 10$^{31}$ cm$^{-2}$ s$^{-1}$ for a 250 GeV
electron beam, assuming proton beam cooling
with
sufficiently small cooling times. The use of a dynamic focusing scheme could
provide
 a potential  
luminosity upgrade by a factor 3-4, and further possible
improvements are not excluded.
However, for the time being we will take the conservative number above as a
guideline, which incidently is very close to the present luminosity 
a HERA. 
Some aspects  of the present physics program 
at HERA are  investigated.
An integrated luminosity of 100 pb$^{-1}$ per year is assumed for this study.
Collisions of 820 GeV protons 
with electrons of 250 GeV can be achieved 'relatively easily' 
with a TESLA type of LC. If the full capacitance of the 
 LC is used, i.e. both the $e^+$ and the $e^-$
accelerator, 500 GeV electron beams (giving roughly half of the 
$ep$ luminosity  compared to above) can be provided with the standard LC, and 
800 GeV electron beams with a possible future upgrade
of the LC. We will mainly study the case for a 
250 GeV electron beam in the following, which would represent the 'day one'
type of LC-HERAp collider.

\section{Kinematics and detector issues}
 \begin{figure}[htbp]
\vspace*{13pt}
\begin{center}
\psfig{figure=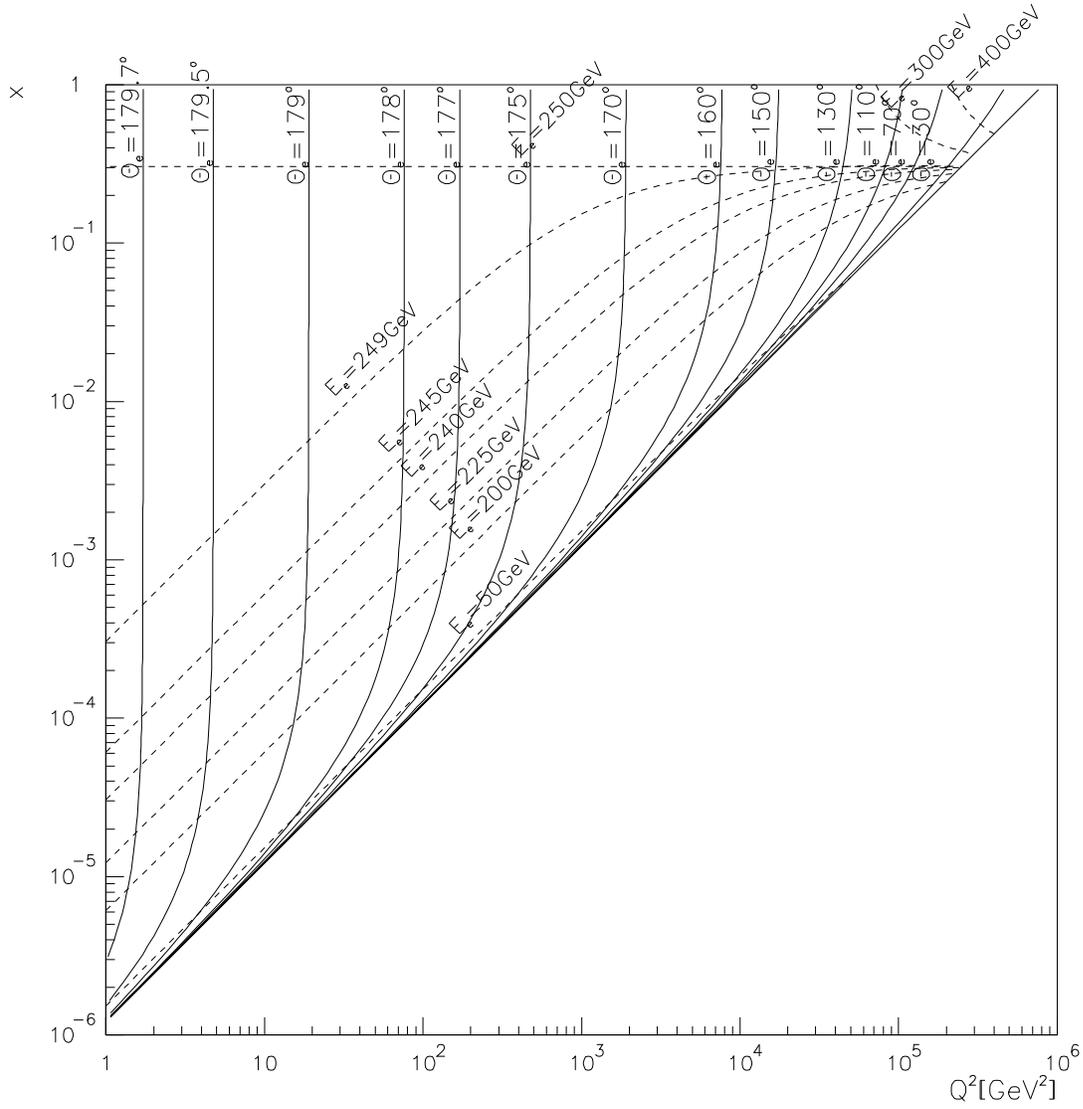,bbllx=0pt,bblly=80pt,bburx=600pt,bbury=750pt,width=17cm}
\caption{Kinematic plane in $x$ and $Q^2$ for collisions with $E_e = 250 $ GeV
and $E_p = 820$ GeV. Iso-angular and iso-energy lines are shown  
 for the scattered electron.}
\label{fig:kinematics1}
\end{center}
\end{figure}

 \begin{figure}[htbp]
\vspace*{13pt}
\begin{center}
\psfig{figure=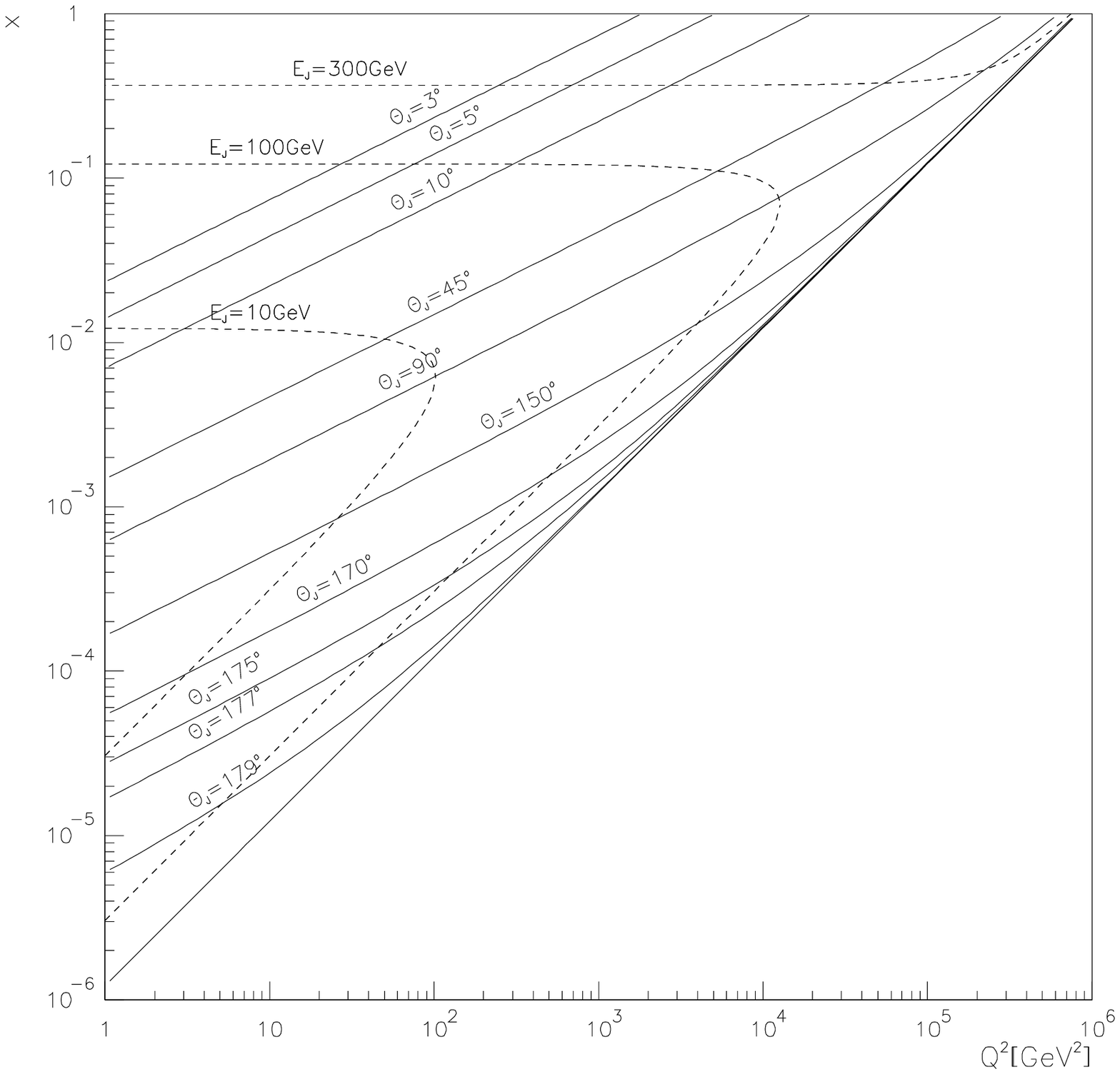,bbllx=0pt,bblly=80pt,bburx=600pt,bbury=750pt,width=17cm}
\caption{Kinematic plane in $x$ and $Q^2$ for collisions with $E_e = 250 $ GeV
and $E_p = 820$ GeV. Iso-angular and iso-energy lines are shown  
 for  the current quark in the  quark parton
model.}
\label{fig:kinematics}
\end{center}
\end{figure}
Figures 1 and 2  show the kinematical plane reachable for a collider with 
 250 GeV electrons and 820 protons. It allows to increase the 
kinematic range by an order of magnitude at large $Q^2$ and small $x$.     
Thus $x$ values down to almost $10^{-6}$ can be reached
in the deep inelastic scattering region ($Q^2> 1$ GeV$^2$). 
Ultimately, collisions of 800 GeV  electrons on 1 TeV beams of protons
will lead to an increase of a 
factor of  about 36 of the CMS energy squared, $s$,  compared to HERA.
In Fig.~1 also the iso-angular and iso-energy 
lines are shown for the scattered
electron. The angle $\theta_e$ is defined in the 
 HERA or LC-HERAp laboratory system to the direction of the proton beam.
Electrons from interactions with $Q^2$ smaller than 100 GeV$^2$ 
 have a small scattering angle
(i.e.~$\theta_e$ close to 180 degrees). Hence to access the low-$x$ 
         low-$Q^2$ region detectors at very low angles, integrated
with the machine elements,  are required. Such compact detectors could be 
built, based on similar techniques now proposed or used in the 
H1~\cite{H1} and ZEUS~\cite{Zeus} experiments at HERA.
For H1  small compact detectors,
called VLQ~\cite{VLQ} detectors,  will be installed during the 97-98 machine
shutdown. These detectors are based on a GaAs tracker and 
a compact Tungsten/Scintillator 
 calorimeter. 
%Note that there is no problem of synchrotron radiation
%in this accelerator configuration at the interaction point.

 \begin{figure}[htbp]
\vspace*{13pt}
\begin{center}
\psfig{figure=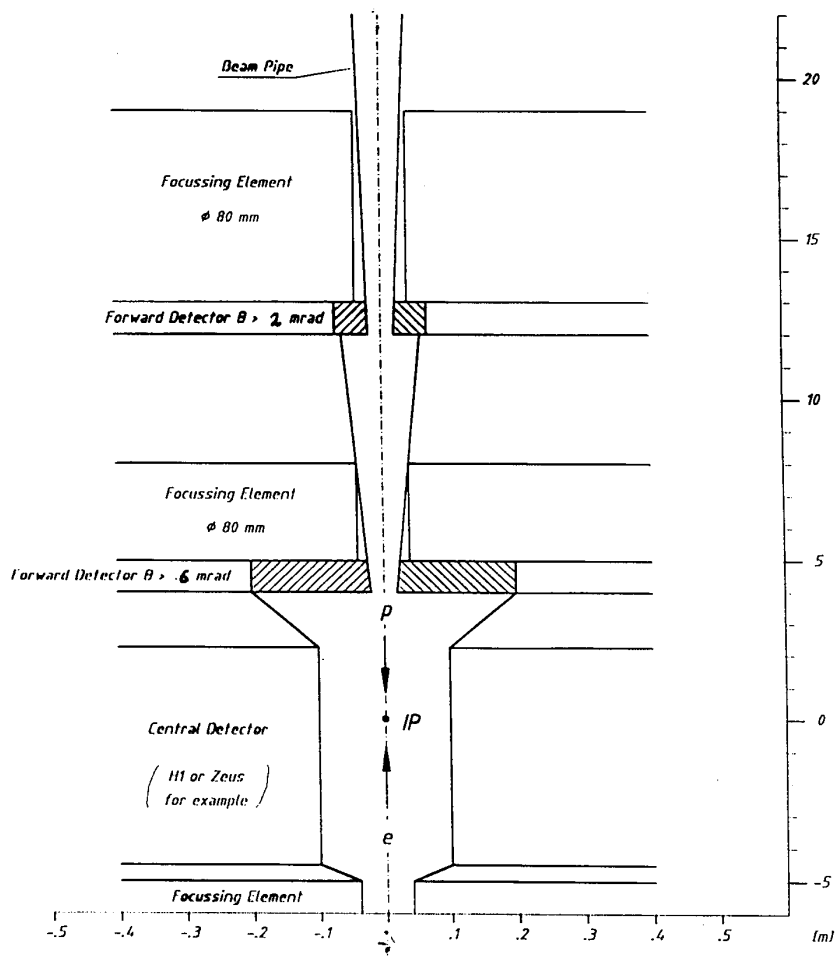,bbllx=0pt,bblly=50pt,bburx=600pt,bbury=750pt,height=15cm}
\caption{A possible layout of a detector at LC-HERAp, consisting of a 
central detector and several compact detectors close to the
beampipe, 
 taking into account the (preliminary) requirements
of the machine. In this picture the proton direction is from top to bottom.}
\label{fig:detector}
\end{center}
\end{figure}

The detector assumed  for these studies consists of
a central detector, taken to be an existing one, H1 or ZEUS, equipped 
with small angle electron detectors, as shown in Fig.~3.
Such a detector would allow to measure 
scattered electrons with angles down to 179.7 degrees.
Note that for this cost-efficient solution one of the experiments would
still need to be moved 
to the  West Hall at the DESY site, tentatively foreseen for the 
location of the $ep$ interaction point, and e.g. calorimeter electronics
needs to be modified for the much larger energies at LC-HERAp.
Note also that if the interaction point is in the West Hall the protons 
need to be accelerated in HERAp in the opposite direction compared 
to the one presently at HERA.
A further potential weakness of re-using existing HERA detectors
 is their asymmetry: the backward regions 
(large $\theta_e$ region) are much poorer equipped for   measuring hadronic
final states (charge particles, energy flows, jets...) compared
to the forward ones. The impact of 
this asymmetry will be discussed below.
Fig.~2 shows the iso-angle and iso-energy lines for the 
current quark jet (for QPM events). 
The down left corner (low-$x$, low-$Q^2$ region)
       shows that there can be a large energy  flow in the backward direction.

\section{Structure functions}

One of the flagship measurements at HERA is  the measurement 
of structure functions in a large kinematic region.
An important issue here is, apart from event statistics,  the quality of the 
 reconstruction of the kinematics of the scattering process.

\begin{figure}
\begin{center}
\psfig{figure=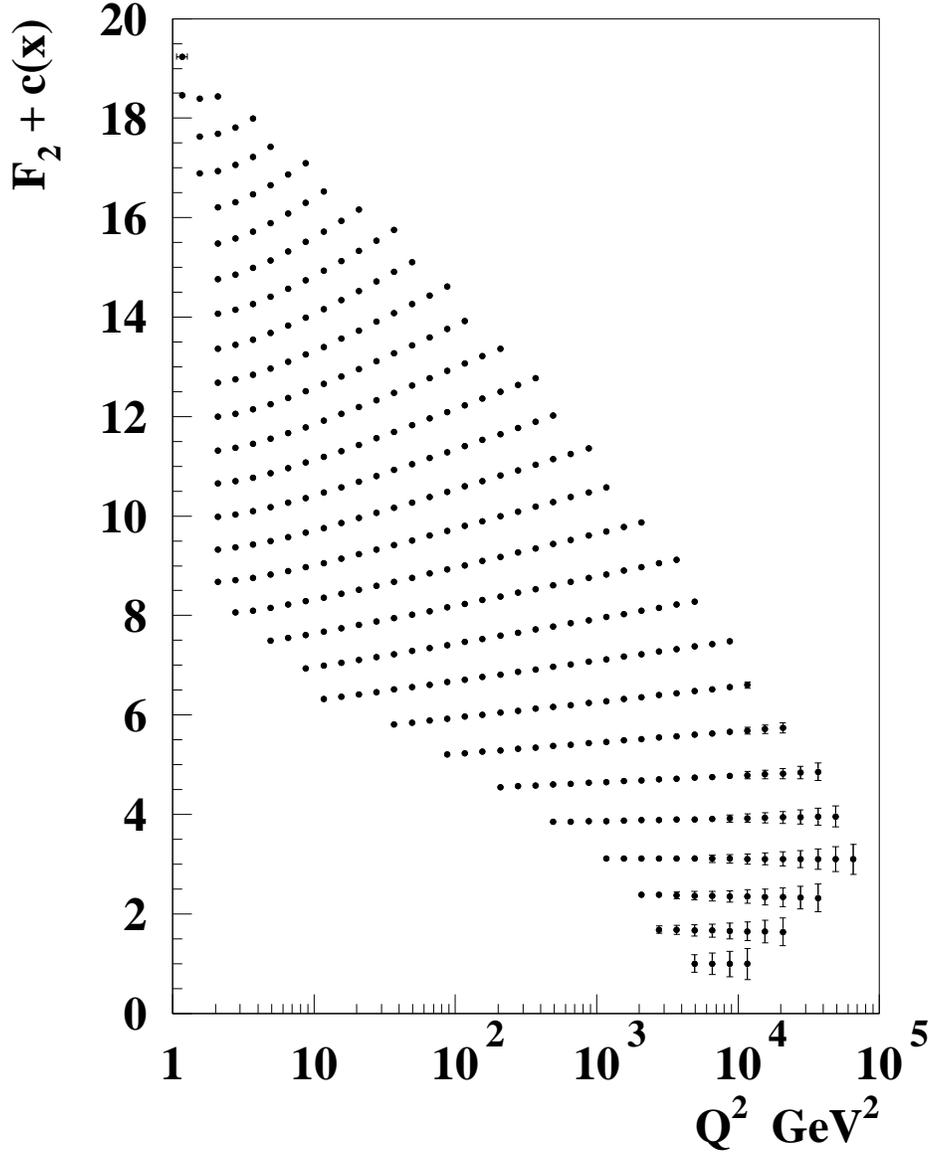,bbllx=0pt,bblly=100pt,bburx=600pt,bbury=750pt,width=18cm}
\caption{Projected measurements of the structure function $F_2$ for 
100 pb$^{-1}$, assuming  $\theta_e<179.7$, and $y_e< 0.9$, and at least
10 events/bin. The errors are the squared sum of the statistical errors 
(assuming 100\% efficiency for the event detection) and a 2\% systematical 
error. The values of $F_2$ plotted are $F_2 + c(x)$, where $c(x)$
is $0.6\cdot (J+0.6)$. $J$ is the bin number in $x$, with $J= 1, 2, 3, 4, 5
$ corresponding to  $x=0.52, 0.32, 0.20, 0.13, 0.082$ etc. The smallest 
$x$ value is $2.10^{-6}$.}
\label{fig:f2}
\end{center}
\end{figure}

The event kinematics can be reconstructed using the energy of
the scattered lepton $E_{e}'$ and the polar angle $\theta _{e}$
according to the relations:
\begin{equation}
  y_e   =1-\frac{E'_e}{E_e} \sin^{2}\frac {\theta_e} {2}
   \hspace*{2cm}
   Q^2_e = 4E'_eE_e\cos^2\frac{\theta_e}{2}
= \frac{E^{'2}_e \sin^2{\theta_e}}{ 1-y_e}. \label{kinele}
\end{equation}
where $E_e$ is the electron beam energy.
The variable $x$ is then calculated as $x = Q^2/sy$.
 This reconstruction method, based on the
scattered
lepton, is called the electron method. It is best applicable
at large values of $y$, and found to give a good resolution for 
$y > 0.1$ at HERA energies. As $\Delta x/x \sim \Delta E_e/(y E_e) 
\sim 1/(y \sqrt{E_e})$
the higher electron energy of the LC-HERAp collider  allows to cover
a range of a factor 3 lower in $y$ compared to HERA.

The kinematics can also be calculated from 
hadronic final states. An example is the 
 $\Sigma$ method~\cite{bernardi} which is used by 
H1 to determine $F_{2}$ at low $y$.
 This method
 combines the electron and the hadronic measurements by defining:
\begin{equation}
   y_{\Sigma} = \frac{\Sigma}{ \Sigma + E'_e(1-\cos{\theta_e})}
   \hspace*{2cm}
   Q^2_{\Sigma} = \frac{E^{'2}_e \sin^2{\theta_e}}{ 1-y_{\Sigma}},
   \label{kinsig}
\end{equation}
with
\begin{equation}
   \Sigma=\sum_h{(E_h-P_{z,h})}.
\end{equation}
Here   $E_h$ and $P_{z,h}$ are the energy and longitudinal momentum
component
of a particle $h$, the summation is  over all hadronic final
state particles, and  masses are neglected.
The denominator of $y_{\Sigma}$ is equal to twice
the energy of the true incident beam energy,
which differs from the nominal beam energy if  real photons
are  emitted by the incident electron.
 LC-HERAp is also favourable  
for the hadronic measurement of the kinematics, as noise contributions 
at low $y$ get  suppressed by an additional  factor 
of 10 compared to HERA,  due to the higher energy of the scattered electron.
 The real limitation comes from beampipe losses (see Fig.~2):
when the current jet angle $\theta_{jet}$ becomes smaller
than 5-10 degrees, the jet will be partially lost in the beampipe,
preventing a reliable calculation of the event kinematics. However, 
the kinematics allows to cover the region down to $y=0.001$ at low $Q^2$
and  to about $y= 0.01$ at $Q^2 \sim 1000$ GeV$^2$.
A potential problem is the loss of hadrons in the electron direction:
the jet angle is close to the beampipe at very small $x$. 
Monte Carlo studies show however
that only  large $y $ $(y> 0.3)$, low-$Q^2$ points are seriously
affected. This is a region which is
 well measured by the electron method,
hence no extra hadronic calorimetry in the electron direction is required 
for structure function measurements. 
Note that the measurements will also have a large overlap with the 
present HERA measurements in the $x- Q^2$ plane, due to the low
$y$ reach at LC-HERAp

An integrated
 luminosity of 100 pb$^{-1}$ would lead to more than $10^{7}$ events/bin
around  $Q^2 \sim $ few
GeV$^2$ at low $x$, more than sufficient for precision 
measurements of structure functions. Above $Q^2 = 10,000$  GeV$^2$
only a few thousand events are expected.
Fig.~4 shows a projection of the structure function measurements for 
an integrated luminosity of 100 pb$^{-1}$ with error bars including 
the statistical errors and a systematical error of 2\%,
added quadratically. Data 
points are shown for which  at least 
10 events/bin are expected. 
The H1 $F_2$ parametrization as given
in~\cite{F294} is used for the extrapolation.
In $Q^2$ 8 bins per decade are taken, while in $x$ 5 bins per decade 
are assumed. 
With 100 pb$^{-1}$, the structure function $F_2$ can be measured up
to 
$Q^2=$ 10$^5$ GeV$^2$, and beyond if a coarser bin size is chosen.

The extended range will allow further 
QCD studies of the structure function and 
the extraction of the gluon density $xg(x)$ from scaling violations.
Hence the gluon density for $x$ values
 down to 10$^{-5}$--10$^{-6}$ 
will be extracted. This will extend considerably the 
program of studies on low-$x$ parton dynamics, in particular studies 
of BFKL effects in the parton densities and search for 
saturation effects. To that end it is of interest to consider the naive
saturation limit, given by Mueller in~\cite{mueller}, where from
geometrical arguments one finds that  the full available 
(transverse) space in the proton 
is filled when the limit
$xg(x) \simeq 6 Q^2$ is reached. Hence for a $Q^2$ 
value of 5 GeV$^2$ the saturation limit would be reached
for the value 
$xg(x) = 30$. HERA measures presently $xg(x) \sim 15$ at $x = 10^{-4}$,
suggesting that the  limit would be reached in the 
region $x \sim 10^{-5} - 10^{-6}$, within the reach of the LC-HERAp.

\section{Hadronic final states and photoproduction}
\begin{figure}
\begin{center}
\psfig{figure=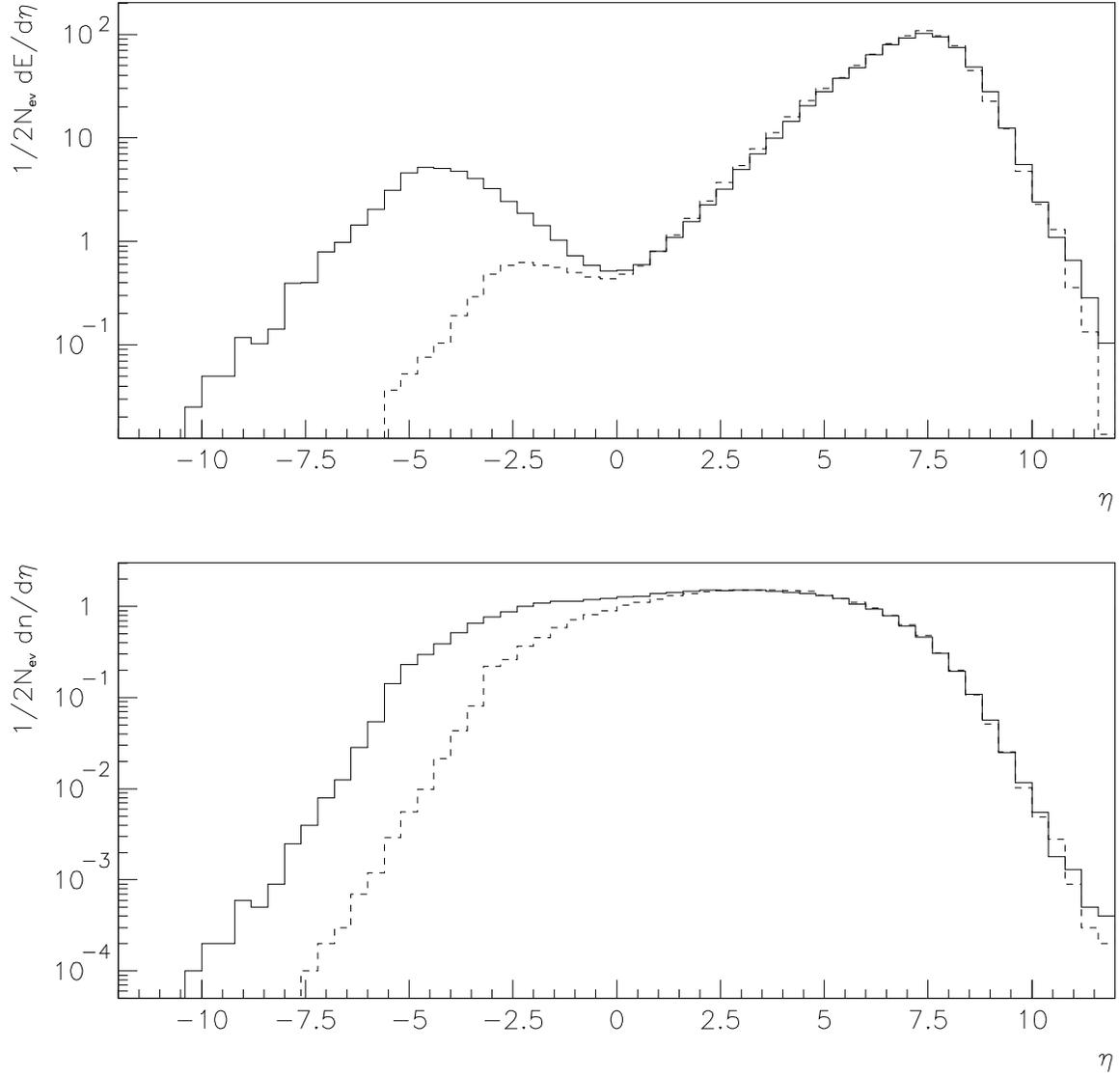,bbllx=0pt,bblly=100pt,bburx=600pt,bbury=750pt,height=20cm}
\caption{Energy (top) and 
 particle (bottom) flow as function of pseudo-rapidity,
for DIS events with $Q^2>5$ GeV$^2$
at HERA (dashed lines) and LC-HERAp (full lines), in the laboratory frame.
Negative pseudo-rapidities correspond to the current quark (or virtual photon)
 direction.
}
\label{fig:eflow}
\end{center}
\end{figure}

Due to the large coverage of the final state phase space by the detectors,
 detailed analyses  can be performed. These analyses of final states
have been an important contribution to the 
 success of the present HERA physics program.
Fig.~5 show the energy and particle flow for HERA and LC-HERAp
as predicted by the Monte Carlo program LEPTO~\cite{lepto}, as function
of pseudo-rapidity $\eta = - \log\tan\theta/2$. Positive values are in the 
proton direction. The events are selected with $Q^2> 5 $ GeV$^2$.
Fig.~5 shows that in order 
to obtain a full detection of the hadronic final state
in the current quark region, hadronic coverage will be necessary
down  to small
angles. For events with 
 $Q^2 > 100 $ GeV$^2$ it suffices to detect hadrons down 
to $\eta =  -5$ to have full current quark hemisphere coverage.

A hot topic presently at HERA is the study of large rapidity gap events, 
generally identified with diffractive events, constituting
 approximately 10\% of the 
total deep inelastic scattering 
 cross section~\cite{diff}. If the system $M_x$, i.e. the dissociated hadronic  
system at the photon side of the phase space, can be detected 
with the future detectors (with the same requirement
as above, namely   detecting the full current
quark region), the increased $W^2$ (= the $\gamma^*p$ invariant
mass squared) of  interactions 
at LC-HERAp, compared to HERA,
allows to reach a larger $M_x$ and $\beta$ region, when 
applying the present selection  of $x_{\pomeron}$ for a 
diffractive sample
 (e.g. $x_{\pomeron} < 0.01$).
Here:
\begin{eqnarray*}
  x_{\pomeron} = \frac{Q^2 + M_x^2}{Q^2 + W^2} \hspace{1.0cm} 
  \beta = \frac{Q^2}{Q^2 + M_x^2}.
\end{eqnarray*}
where in the infinite momentum frame $\beta$ acquires the interpretation
of the Bjorken-$x$ scaling variable of the parton probed in the 
diffractive exchange.
E.g. for $y$ = 0.33 and $Q^2 = 10 $ GeV$^{2}$ we have $M_x^{max} = 15$ GeV 
and $\beta^{min} = 0.03$ at HERA while $M_x^{max} = 50$ GeV and $\beta^{min}
= 0.003$ at LC-HERAp. The larger $M_x$ region covered
 yields more phase space for e.g studies of
jets with large transverse momentum $E_T$
 in diffractive systems. The region at low $\beta$ is unexplored at present.
Does the diffractive structure function rise at low $\beta$,
similarly to  the 
proton structure function  at low $x$? Questions of this sort
 can 
be addressed at the LC-HERAp.

Heavy flavour studies will  benefit from the energy increase.
Charm and bottom cross sections increase with a factor of about 3 to 4, 
compared to HERA\cite{ali}.
Heavy flavours are dominantly produced via the boson-gluon fusion
diagram in photoproduction. Hence the cross section is directly 
proportional to the gluon density in the proton. The extended kinematical 
range of the LC-HERAp allows to cover $x$ values down to 10$^{-4}$.
The low-$x$ charmed mesons (e.g $D^*$) are however produced in the electron
direction, 
at pseudo-rapidities of typically $-2$ to $-4$. Hence to exploit the  
  sensitivity to the gluon  at the lowest 
$x$ values, charm detection in the backward  region must be foreseen for the
detectors. LC-HERAp will however not become a machine to study top quarks.
 The expected number of top
quarks produced will be of the order of a few tens per year, even at the 
highest
energies.
Clearly the $e^+e^-$ LC itself is the machine 
to study  this topic.

Finally, the presence of electron taggers  for electrons with a very small 
scattering angle, similar to the ones currently used by the 
HERA experiments,
will allow a measurement of the real photoproduction
  cross sections at $W_{\gamma p}
\sim 600$ GeV. Together 
with the present HERA data this gives a substantial lever arm
for 
studies of the energy dependence of total, 
inclusive and exclusive photoproduction cross sections,
 and studies of the interesting questions related to the transition from 
soft to hard physics, a topic presently explored at HERA.

\section{High \boldmath{$Q^2$} region}
\begin{figure}[t]
\begin{center}
\psfig{figure=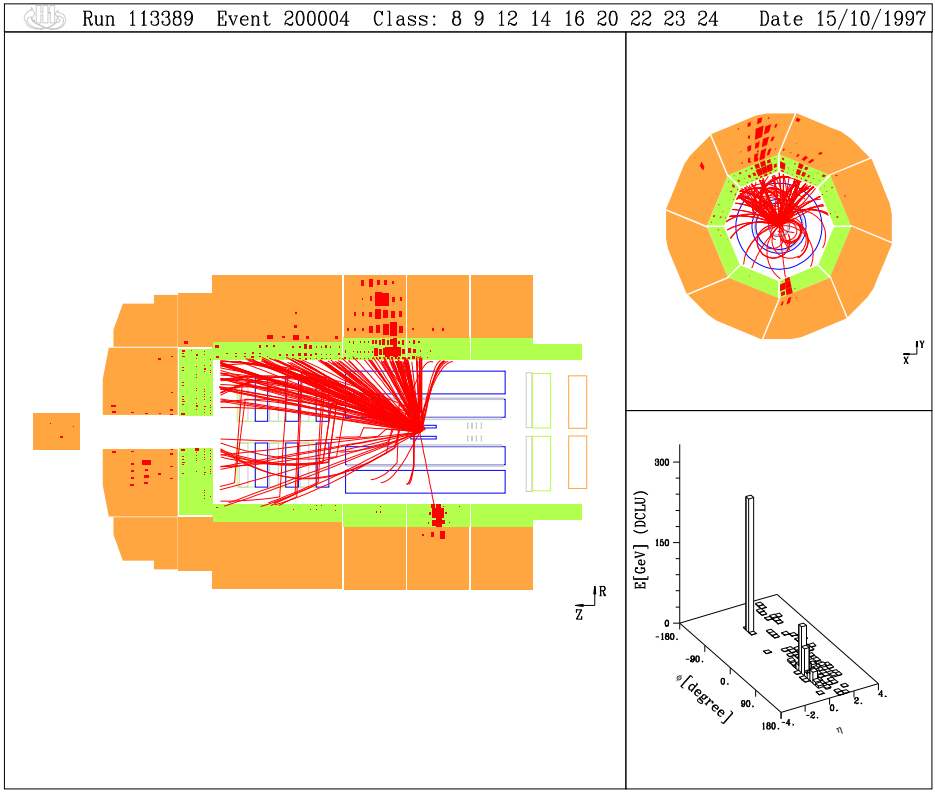,bbllx=0pt,bblly=150pt,bburx=650pt,bbury=650pt,width=12cm,angle=90}
\caption{A fully simulated and reconstructed event in the H1 detector 
for LC-HERAp ($E_e = 250$ GeV and $E_p$ = 820 GeV), with $x = 0.3$ and
$Q^2 = 10^5$ GeV$^2$. Protons enter from the right, electrons from the left.
The scattered electron (bottom) is well isolated from the scattered quark
jet (top). The pictures on the right show the transverse view of the 
central region of the detector and a lego plot of the energy deposits
in the calorimeter.}
\label{fig:event}
\end{center}
\end{figure}
%\begin{figure}[htb]
%\begin{center}
%\psfig{figure=parton.ps,bbllx=-200pt,bblly=100pt,bburx=325pt,bbury=700pt,height=8cm}
%\caption{The partonic cross sections for different colliders}
%\label{fig:gluon_over}
%\end{center}
%\end{figure}
The recently reported  excess of events at high $Q^2$ at HERA~\cite{wagner}
 is a clear physics case which strongly supports the road for
$ep$ collisions at higher energies. For the same 
$(Q^2,y)$  region the cross sections rise
substantially with increasing collision energy, since the proton is  probed
at  relatively lower $x$ values.
Threshold effects are overcome and e.g. in case of contact interactions,
the larger $Q^2$ phase space available allows for  more pronounced effects
to be observed. Hence increasing the energy of the $ep$ collision will
provide vital information on 
the nature of the high-$Q^2$ anomaly, if it exists.

The present H1 and ZEUS detectors are very well tailored to measure and 
analyse high-$Q^2$ events in the LC-HERAp environment, as is shown in Fig.~6.
Here the simulation result of an LC-HERAp 
event with $x= 0.3$ and $Q^2 = 100,000$
GeV$^2$ is shown for the H1 detector. The electron is scattered in the central 
part of the detector (bottom) and quite isolated from the hadronic final state,
hence well measurable.

Table 1 gives the cross section for $Q^2 > Q^2_{min}$ values.
For $Q^2> 10^5$ GeV$^2$, 20 to 50 events per
 100 pb$^{-1}$ are expected, according
to the Standard Model cross sections (using the GRV~\cite{grv}
 structure function and the DJANGO event generator~\cite{django}).
New effects may dramatically change these numbers, but the table
shows that it  would be clearly
advantageous to have  at least a  luminosity 
which is ten times larger than assumed here, 
in order to  study the high $Q^2$ 
region in detail.

\begin{table}
\hfil
%\begin{flushleft}
%\hspace{-4.5cm}
\begin{tabular}{||c|c|c|c|c||}
\hline\hline
$Q^2>Q^2_{min}$ (GeV$^2$)&$10^4$  &$ 2.10^4
$ & $5.10^4$ &$ 10^5$ \\
\hline\hline
$\sigma(e^+p\rightarrow e^+X)$ & 29 & 9 & 1.3 & 0.2  \\
\hline
$\sigma(e^-p\rightarrow e^-X)$ & 36 & 12 & 2.3 & 0.5  \\
\hline\hline
\end{tabular}
\caption{Integrated cross sections for $Q^2>Q^2_{min}$ for $e^+p$ and $e^-p$
neutral current events, in picobarn.}
\end{table}
\section{Summary}
We have discussed some of the physics topics which could be addressed 
at LC-HERAp. Already in its most conservative mode ($E_e$ = 250 GeV,
luminosity of the same order as presently at HERA) a rich physics 
program can emerge at relatively 'low cost', assuming the re-use of
one of the present detectors, equipped with detectors for scattered
electrons with low angles. Structure functions can be measured at lower
$x$, re-opening search for novel low-$x$ phenomena and saturation, 
and at higher $Q^2$. To exploit maximally the continuation of the 
presently successful analysis program on final states, an improved
hadronic coverage in the backward direction is mandatory.
The new  high-$Q^2$ region can be studied rather easily
 with a central detector,
but higher luminosities would be desirable.

\section*{Acknowledgments}
I thank the organizers for providing a well organized and very stimulating 
workshop in Ankara. I would like to thank for R. Brinkmann and R. Felst
for useful discussions.

\end{document}